\documentstyle[12pt]{article}
\begin{document}
\begin{titlepage}
\begin{flushright}
IC/2001/146\\
hep-th/0111118
\end{flushright}
\vspace{10 mm}

\begin{center}
{\Large A Varying-$e$ Brane World Cosmology}

\vspace{5mm}

\end{center}
\vspace{5 mm}

\begin{center}
{\large Donam Youm\footnote{E-mail: youmd@ictp.trieste.it}}

\vspace{3mm}

ICTP, Strada Costiera 11, 34014 Trieste, Italy

\end{center}

\vspace{1cm}

\begin{center}
{\large Abstract}
\end{center}

\noindent

We study a varying electric charge brane world cosmology in the RS2 
model obtained from a varying-speed-of-light brane world cosmology 
by redefining the system of units.  We elaborate conditions under 
which the flatness problem and the cosmological constant problem 
can be resolved by such cosmological model.

\vspace{1cm}
\begin{flushleft}
November, 2001
\end{flushleft}
\end{titlepage}
\newpage

The recent observational evidence \cite{wfc,mur,web} for time-varying 
fine-structure constant $\alpha=e^2/(4\pi\hbar c)$ prompted renewed 
interest in cosmological models where a fundamental constant of nature, 
e.g., the speed of light $c$ or the electric charge $e$, varies with time.  
Variable-Speed-of-Light (VSL) cosmological models \cite{mof1,am} were 
proposed also as an alternative to inflation \cite{gut,lin,als} for 
resolving \cite{mof1,am,bar1,bar2,bar3,mof2,cm1,bar4,cm2,blm,cm3,cm4} 
the cosmological problems of the Standard Big Bang model.  The original 
VSL models by Moffat \cite{mof1} and by Albrecht and Magueijo \cite{am} 
assume that the speed of light, a constant in the action, varies with time 
during an early period of cosmic evolution, thereby the Lorentz symmetry 
becoming explicitly broken.  The bimetric models, later proposed by Clayton 
and Moffat \cite{cm1,cm2,cm3,cm4}, provide a mechanism by which the speed 
of light can vary with time in a diffeomorphism invariant manner without 
explicitly breaking the Lorentz symmetry.  (See also Ref. \cite{dru} for 
an independent development.)  
Alternatively, a varying-$\alpha$ cosmological model where the electric 
charge $e$ varies with time in the manner described by the varying-$\alpha$ 
theory of Bekenstein \cite{bek} was also studied \cite{sbm} in an attempt 
to give theoretical explanation for currently observed time-varying $\alpha$.  

Previously, we studied \cite{youm,youm1,youm2} the VSL cosmologies in the 
Randall-Sundrum (RS) scenarios \cite{rs1,rs2}, following the approaches of 
the VSL models with varying fundamental constant and the bimetric model of 
Clayton and Moffat.  We studied conditions under which the cosmological 
problems can be resolved by such VSL brane world cosmologies.  In particular, 
it was shown that the cosmological constant problem in the RS brane world 
cosmologies can be resolved by the VSL theories.  So, the VSL models can be 
used to bring the corrections to the fine-tuned brane tensions under 
control.  The VSL models in the brane world scenarios are of interest, also 
because of the recent works \cite{kal1,kir,kal2,chu,ale,ish,ckr,csa} 
indicating the Lorentz violation in the brane world scenarios.  We also 
studied \cite{youm3} the brane world cosmology in the RS model where the 
electric charge varies with time in the manner described by the 
varying-$\alpha$ theory of Bekenstein \cite{bek}.  

In this paper, we study the varying-$e$ brane world cosmology which is 
related to the VSL brane world cosmology through the change of system 
of units.  We consider the Randall-Sundrum (RS) brane world model with 
one positive tension brane and infinite extra dimensions, i.e., the 
RS2 model \cite{rs2}.  We summarize the VSL cosmology in the 
RS2 model studied in Ref. \cite{youm}.  Then, we redefine the system 
of units to map the VSL brane world cosmology to the dual varying-$e$ 
brane world cosmology.  We find that the flatness and the cosmological 
constant problems can be resolved by such varying-$e$ cosmological model, 
provided the dielectric field varies rapidly enough.  

We begin by discussing the VSL brane world cosmology studied in Ref. 
\cite{youm}.  If the speed of light $c$ is variable, then the Lorentz 
invariance becomes explicitly broken.  Therefore, it is postulated in VSL 
models that there exists a preferred Lorentz frame in which the laws of 
physics simplify.  In such preferred frame, the action is assumed to be 
given by the standard action with a constant $c$ is replaced by a field 
$c(x^{\mu})$, the so-called {\it principle of minimal coupling}.  In a 
preferred Lorentz frame, the action for the VSL cosmology in the RS2 model 
is given by
\begin{equation}
S=\int d^5x\left[\sqrt{-\hat{g}}\left({\psi\over{16\pi G_5}}{\cal R}-\Lambda
\right)+{\cal L}_{\psi}\right]+\int d^4x\sqrt{-g}\left[{\cal L}_{\rm mat}
-\sigma\right],
\label{rs2act}
\end{equation}
where $\sigma$ is the tension of the 3-brane (assumed to be located at the 
origin $y=0$ of the extra spatial coordinate $y$), a scalar field $\psi(x^M)
\equiv c^4(x^M)$ is defined out of varying speed of light $c(x^M)$, the 
Lagrangian ${\cal L}_{\psi}$ for $\psi$ controls the dynamics of $\psi$, 
and ${\cal L}_{\rm mat}$ is the Lagrangian density of matter fields 
confined on the brane.  It is required that ${\cal L}_{\psi}$ should be 
explicitly independent of the other fields, including the metric, so that 
the principle of minimal coupling continues to hold for the equations of 
motion.   

Since $\psi$ is assumed to be minimally coupled, the Einstein's equations 
take the conventional forms with the constant $c$ replaced by a field $c(x^M)$:
\begin{equation}
{\cal G}_{MN}={{8\pi G_5}\over\psi}{\cal T}_{MN},
\label{eineqs}
\end{equation}
with the energy-momentum tensor given by
\begin{equation}
{\cal T}_{MN}=-\hat{g}_{MN}\Lambda+\delta^{\mu}_M\delta^{\nu}_N\left(
{\cal T}^{\rm mat}_{\mu\nu}-g_{\mu\nu}\sigma\right){\sqrt{-g}\over
\sqrt{-\hat{g}}}\delta(y).
\label{rs2emtens}
\end{equation}
Here, the energy-momentum tensor ${\cal T}^{\rm mat}_{\mu\nu}=-{2\over
\sqrt{-g}}{{\delta(\sqrt{-g}{\cal L}_{\rm mat})}\over{\delta g^{\mu\nu}}}$ 
for the brane matter fields in the comoving frame has the usual perfect 
fluid form:
\begin{equation}
{\cal T}^{\rm mat\ \mu}_{\ \ \ \ \ \ \nu}={\rm diag}\left(-\varrho c^2,
\wp,\wp,\wp\right),
\label{pfem}
\end{equation}
where $\varrho$ and $\wp$ are the mass density and the pressure of the 
brane matter fields.  

The general bulk metric Ansatz for the expanding brane universe where the 
principles of homogeneity and isotropy in the three-dimensional subspace are 
satisfied is 
\begin{equation}
\hat{g}_{MN}dx^Mdx^N=-n^2(t,y)c^2dt^2+a^2(t,y)\gamma_{ij}dx^idx^j+b^2(t,y)dy^2,
\label{bulkmet}
\end{equation}
where $\gamma_{ij}$ is the metric for the maximally symmetric 
three-dimensional space given by
\begin{equation}
\gamma_{ij}dx^idx^j=\left(1+{k\over 4}\delta_{mn}x^mx^n\right)^{-2}\delta_{ij}
dx^idx^j={{dr^2}\over{1-kr^2}}+r^2(d\theta^2+\sin^2\theta d\psi^2),
\label{gammet}
\end{equation}
with $k=-1,0,1$ respectively for the three-dimensional spaces with the 
negative, zero and positive spatial curvatures.  
Then, the Einstein's equations (\ref{eineqs}) take the forms:
\begin{equation}
{3\over {c^2n^2}}{\dot{a}\over a}\left({\dot{a}\over a}+{\dot{b}\over b}
\right)-{3\over b^2}\left[{a^{\prime}\over a}\left({a^{\prime}\over a}-
{b^{\prime}\over b}\right)+{a^{\prime\prime}\over a}\right]+{{3k}\over a^2}=
{{8\pi G_5}\over c^4}\left[\Lambda+(\sigma+\varrho c^2){{\delta(y)}\over b}
\right],
\label{ttein}
\end{equation}
\begin{eqnarray}
{1\over b^2}\left[{a^{\prime}\over a}\left(2{n^{\prime}\over n}+{a^{\prime}
\over a}\right)-{b^{\prime}\over b}\left({n^{\prime}\over n}+2{a^{\prime}
\over a}\right)+2{a^{\prime\prime}\over a}+{n^{\prime\prime}\over n}\right]+
\ \ \ \ \ \ \ \ \ \ \ \ \ \ \ \ \ \ \ \ \ \ \ \ \ \ \ \ \ \ \ \ \ \ \ \ \ \
\ \ \
\cr
{1\over {c^2n^2}}\left[{\dot{a}\over a}\left(2{\dot{n}\over n}
-{\dot{a}\over a}\right)+{\dot{b}\over b}\left({\dot{n}\over n}
-2{\dot{a}\over a}\right)-2{\ddot{a}\over a}-{\ddot{b}\over b}\right]
-{k\over a^2}=
-{{8\pi G_5}\over c^4}\left[\Lambda+(\sigma-\wp){{\delta(y)}\over b}\right],
\label{iiein}
\end{eqnarray}
\begin{equation}
{n^{\prime}\over n}{\dot{a}\over a}+{a^{\prime}\over a}{\dot{b}\over b}
-{\dot{a}^{\prime}\over a}=0,
\label{tyein}
\end{equation}
\begin{equation}
{3\over b^2}{a^{\prime}\over a}\left({a^{\prime}\over a}+{n^{\prime}\over n}
\right)-{3\over{c^2n^2}}\left[{\dot{a}\over a}\left({\dot{a}\over a}-{\dot{n}
\over n}\right)+{\ddot{a}\over a}\right]-{{3k}\over
a^2}=-{{8\pi G_5}\over c^4}\Lambda,
\label{yyein}
\end{equation} 
where the overdot and the prime respectively denote derivatives w.r.t. $t$ 
and $y$.  Although the metric components are continuous everywhere, their 
derivatives w.r.t. $y$ are discontinuous at $y=0$ due to the $\delta$-function 
like brane source there.  The following boundary conditions on the first 
derivatives at $y=0$ are determined by Eqs. (\ref{ttein},\ref{iiein}):  
\begin{equation}
{{[a^{\prime}]_0}\over{a_0b_0}}=-{{8\pi G_5}\over{3c^4}}(\sigma+\varrho c^2),
\label{bc1}
\end{equation}
\begin{equation}
{{[n^{\prime}]_0}\over{n_0b_0}}=-{{8\pi G_5}\over{3c^4}}(\sigma-3\wp-2\varrho 
c^2),
\label{bc2}
\end{equation}
where the subscript 0 denotes quantities evaluated at $y=0$, e.g., $a_0(t)
\equiv a(t,0)$, and $[F]_0\equiv F(0^+)-F(0^-)$ denotes the {\it jump} of 
$F(y)$ across $y=0$.  

The effective four-dimensional Friedmann equations on the 3-brane can be 
obtained \cite{bdl} by taking the jumps and the mean values of the 
five-dimensional Einstein's equations (\ref{ttein}-\ref{yyein}) across $y=0$ 
and then applying the boundary conditions (\ref{bc1},\ref{bc2}).  
We assume that the radius of the extra space is stable, i.e., $\dot{b}=0$, and 
the $y$-coordinate is defined so that $b=1$.  The resulting effective 
Friedmann equations have the forms:
\begin{equation}
\left({\dot{a}_0\over a_0}\right)^2={{16\pi^2G^2_5}\over{9c^6}}
(\varrho^2c^4+2\sigma\varrho c^2)+{{{\cal C}c^2}\over 
a^4_0}+{{4\pi G_5}\over{3c^2}}\left(\Lambda+
{{4\pi G_5}\over{3c^4}}\sigma^2\right)-{{kc^2}\over a^2_0},
\label{effred1}
\end{equation}
\begin{equation}
{\ddot{a}_0\over a_0}=-{{16\pi^2G^2_5}\over{9c^6}}(2\varrho^2c^4
+\sigma\varrho c^2+3\sigma\wp+3\wp\varrho c^2)-{{{\cal C}c^2}\over 
a^4_0}+{{4\pi G_5}\over{3c^2}}\left(\Lambda+
{{4\pi G_5}\over{3c^4}}\sigma^2\right),
\label{effred2}
\end{equation}
where ${\cal C}$ is an integration constant.  The ${\cal C}$-term (called 
``dark radiation'' term) originates from the Weyl tensor of the bulk and 
thus describes the backreaction of the bulk gravitational degress of freedom 
on the brane \cite{bdl,bdel,ftw,muk,ida}.  

In the limit of $\sigma\gg\varrho c^2, \wp$ \cite{cgk,cgs} and with an 
assumption of the fine-tuned brane tension $\sigma=\sqrt{-{{3c^4}\over{4\pi 
G_5}}\Lambda}$, the effective Fredmann equations (\ref{effred1},\ref{effred2}) 
take the following forms of the conventional cosmology:
\begin{equation}
\left({\dot{a}_0\over a_0}\right)^2={{8\pi G_4}\over 3}\varrho+{{{\cal C}
c^2}\over a^4_0}-{{kc^2}\over a^2_0},
\label{cvfred1}
\end{equation}
\begin{equation}
{\ddot{a}_0\over a_0}=-{{4\pi G_4}\over 3}(\varrho+3{\wp\over c^2})
-{{{\cal C}c^2}\over a^4_0},
\label{cvfred2}
\end{equation}
where the effective four-dimensional Newton's constant given by
\begin{equation}
G_4={{8\pi G^2_5\sigma}\over{3c^4}}.
\label{effg4}
\end{equation}
From these effective Friedmann equations, we obtain the following generalized 
energy conservation equation:
\begin{equation}
\dot{\varrho}+3{\dot{a}_0\over a_0}\left(\varrho+{\wp\over c^2}\right)=
-{\dot{G}_4\over G_4}\varrho+{{3kc\dot{c}}\over{4\pi G_4a^2_0}}-
{{3{\cal C}c\dot{c}}\over{4\pi G_4a^4_0}}.
\label{encnseq}
\end{equation}
So, while $G_4$ or $c$ varies with time, the matter is created on or taken 
out of the brane universe.  

We now map the VSL brane world cosmology discussed in the above to the dual 
varying-$e$ brane world cosmology by changing the system of units.  
As was pointed out in Ref. \cite{am}, it makes sense only to talk about 
constancy or variability of dimensionless ratios of dimensionful quantities, 
since the measured values of dimensionful quantities are actually the ratios 
to some standard units, which may vary with time.   Depending on a choice of 
units, we can regard time-variation of a dimensionless ratio as being due to 
any subset of dimensionful quantities forming the dimensionless ratio.  For 
example, we can regard the time-variation of $\alpha=e^2/(4\pi\hbar c)$ as 
being due to the time variation of either $c$ (and $\hbar$) or $e$, depending 
on the choice of units.  Furthermore, it is always possible to redefine 
system of units such that a given model is mapped to the dual model where 
a different subset of dimensionful quantities varies with time.  
We redefine the system of units such that the speed of light remains 
constant and the electric charge varies with time.  Quantities in the 
system of units where the speed of light $c$ varies with time and the 
electric charge $e$ remains constant are denoted without hat.  Those in 
the units where the electric charge varies with time and the speed of 
light remains constant are denoted with hat.  We relate these two 
systems of units in the following way (Cf. Ref. \cite{bar2}):
\begin{equation}
c^2dt=\hat{c}^2d\hat{t},\ \ \ \ \ \ \ \ \ \ 
cdx=\hat{c}d\hat{x},\ \ \ \ \ \ \ \ \ \ 
{{dE}\over c^3}={{d\hat{E}}\over\hat{c}^3}.
\label{relunts}
\end{equation}
To find the relations between dimensionful quantities in the two 
systems of units, we consider the following dimensionless ratios of 
the dimensionful quantities:
\begin{equation}
{{cdt}\over{dx}}={{\hat{c}d\hat{t}}\over{d\hat{x}}},\ \ \ \ \ 
{\hbar\over{dEdt}}={\hat{\hbar}\over{d\hat{E}d\hat{t}}},\ \ \ \ \ 
{{G_4dE}\over{c^4dx}}={{\hat{G}_4d\hat{E}}\over{\hat{c}^4d\hat{x}}},
\ \ \ \ \ 
{e^2\over{dEdx}}={\hat{e}^2\over{d\hat{E}d\hat{x}}},
\label{dmlsrts}
\end{equation}
which take the same values regardless of the system of units chosen.  
Here, the time-varying speed of light $c$ and the constant speed of light 
$\hat{c}=c_0$ are assumed to be related by a function $\varepsilon(t)$ of 
time as $c=\hat{c}\varepsilon$.   In the system of units without hat, $\hbar
\propto c\propto 1/\sqrt{\alpha}$, so $\hbar=\hat{\hbar}\varepsilon$.  The 
following relations among the remaining dimensionful quantities in the two 
systems of units can be obtained from Eqs. (\ref{relunts},\ref{dmlsrts}):
\begin{equation}
\hat{e}=e/\varepsilon,\ \ \ \ \ \ \ \ \ 
\hat{G}_4=G_4,
\label{parel}
\end{equation}
from which we see that $\varepsilon$ can be interpreted as the vacuum 
dielectric field.  Since $c=\hat{c}\varepsilon$, Eq. (\ref{relunts}) is 
equivalent to
\begin{equation}
d\hat{t}=\varepsilon^2dt,\ \ \ \ \ 
d\hat{x}=\varepsilon dx,\ \ \ \ \ 
d\hat{E}=dE/\varepsilon^3.
\label{tranrel}
\end{equation}
By using this transformation, we can find relations among measurements 
of any dimensionful quantities in the two systems of units.  In 
particular, the mass densities, the pressures and the brane tensions 
in the two systems of units are related as
\begin{equation}
\hat{\varrho}=\varrho/\varepsilon^4,\ \ \ \ \ \ 
\hat{\wp}=\wp/\varepsilon^6,\ \ \ \ \ \ \ 
\hat{\sigma}=\sigma/\varepsilon^6.
\label{dimqrels}
\end{equation}
From Eqs. (\ref{parel},\ref{dimqrels}), we see that the 
five-dimensional Newton's constants in the two systems of units are 
relationed as $\hat{G}_5=G_5\varepsilon$.

In the new system of units, in which the speed of light remains constant, 
the bulk metric (\ref{bulkmet}) takes the form:
\begin{equation}
d\hat{s}^2=\varepsilon^2ds^2=-\hat{n}^2\hat{c}^2d\hat{t}^2+\hat{a}^2
\hat{\gamma}_{ij}d\hat{x}^id\hat{x}^j+\hat{b}^2d\hat{y}^2,
\label{nwblkmet}
\end{equation}
where $\hat{n}=n$, $\hat{a}=\varepsilon a$, $\hat{b}=b$, and
\begin{equation}
\hat{\gamma}_{ij}d\hat{x}^id\hat{x}^j=\left(1+{\hat{k}\over 4}
\delta_{mn}\hat{x}^m\hat{x}^n\right)\delta_{ij}d\hat{x}^id\hat{x}^j
={{d\hat{r}^2}\over{1-\hat{k}\hat{r}^2}}+\hat{r}^2\left(d\hat{\theta}^2
+\sin^2\hat{\theta}d\hat{\phi}^2\right).
\label{nwgammet}
\end{equation}
After redefining the units, a spatial coordinate transformation 
is performed so that $\hat{k}=k=0,\pm 1$.  
By applying the above transformations between the two systems of units to 
Eqs. (\ref{effred1},\ref{effred2}), we obtain the following effective 
Friedmann equations in the new system of units:
\begin{equation}
\left({\dot{\hat{a}}_0\over\hat{a}_0}-{\dot{\varepsilon}\over\varepsilon}
\right)^2=
{{16\pi^2\hat{G}^2_5}\over{9c^6_0}}(\hat{\varrho}^2c^4_0+2\hat{\sigma}
\hat{\varrho}c^2_0)+{{\hat{\cal C}c^2_0}\over\hat{a}^4_0}
+{{4\pi\hat{G}_5}\over{3c^2_0}}\left(\hat{\Lambda}+{{4\pi\hat{G}_5}
\over{3c^4_0}}\hat{\sigma}^2\right)-{{\hat{k}c^2_0}\over\hat{a}^2_0},
\label{nfrd1}
\end{equation}
\begin{equation}
{\ddot{\hat{a}}_0\over\hat{a}_0}-{\ddot{\varepsilon}\over\varepsilon}=
-{{16\pi^2\hat{G}^2_5}\over{9c^6_0}}(2\hat{\varrho}^2c^4_0
+\hat{\sigma}\hat{\varrho}c^2_0+3\hat{\sigma}\hat{\wp}+3\hat{\wp}\hat{\varrho}
c^2_0)-{{\hat{\cal C}c^2_0}\over\hat{a}^4_0}+{{4\pi\hat{G}_5}\over{3c^2_0}}
\left(\hat{\Lambda}+{{4\pi\hat{G}_5}\over{3c^4_0}}\hat{\sigma}^2\right),
\label{nfrd2}
\end{equation}
where $\hat{\cal C}=\varepsilon^2{\cal C}$, $\hat{\Lambda}=
\Lambda/\varepsilon$ and the overdot from now on denotes derivative 
w.r.t. $\hat{t}$.  
In the limit of $\hat{\sigma}\gg\hat{\varrho}c^2_0, \hat{\wp}$, these 
effective Friedmann equations take the forms:
\begin{equation}
\left({\dot{\hat{a}}_0\over\hat{a}_0}-{\dot{\varepsilon}\over\varepsilon}
\right)^2={{8\pi\hat{G}_4}\over 3}\hat{\varrho}+{{\hat{\cal C}c^2_0}\over
\hat{a}^4_0}-{{\hat{k}c^2_0}\over\hat{a}^2_0},
\label{lnfrd1}
\end{equation}
\begin{equation}
{\ddot{\hat{a}}_0\over\hat{a}_0}-{\ddot{\varepsilon}\over\varepsilon}=
-{{4\pi\hat{G}_4}\over 3}(\hat{\varrho}+3{\hat{\wp}\over c^2_0})-
{{\hat{\cal C}c^2_0}\over\hat{a}^4_0},
\label{lnfrd2}
\end{equation}
where $\hat{G}_4=8\pi\hat{G}^2_5\hat{\sigma}/3c^4_0$.
The energy conservation equation (\ref{encnseq}) is transformed to
\begin{equation}
\dot{\hat{\varrho}}+3{\dot{\hat{a}}_0\over\hat{a}_0}\left(\hat{\varrho}+
{\hat{\wp}\over c^2_0}\right)=\left(3{\hat{\wp}\over c^2_0}-\hat{\varrho}+
{{3\hat{k}c^2_0}\over{4\pi\hat{G}_4\hat{a}^2_0\varepsilon^2}}-{{3\hat{\cal C}
c^2_0}\over{4\pi\hat{G}_4\hat{a}^4_0\varepsilon^2}}\right){\dot{\varepsilon}
\over\varepsilon}-{\dot{\hat{G}}_4\over\hat{G}_4}\hat{\varrho},
\label{neceq}
\end{equation}
which can be obtained also by taking the $\hat{t}$-derivatives of Eqs. 
(\ref{lnfrd1},\ref{lnfrd2}).  

We now discuss the resolution of cosmological problems by our dual 
varying-$e$ brane world cosmological model.  
First, we consider the resolution of the flatness problem.  
From Eq. (\ref{lnfrd1}) we see that the critical 
mass density $\hat{\varrho}_c$, defined as the mass density of a flat 
universe ($\hat{k}=0$) for a given Hubble parameter 
$\dot{\hat{a}}_0/\hat{a}_0$, is given by
\begin{equation}
\hat{\varrho}_c={3\over{8\pi\hat{G}_4}}\left[\left({\dot{\hat{a}}_0\over 
\hat{a}_0}-{\dot{\varepsilon}\over\varepsilon}\right)^2-{{\hat{C}c^2_0}\over 
\hat{a}^4_0}\right].
\label{critdens}
\end{equation}
We measure the deviation of the mass density $\hat{\varrho}$ of the universe 
from the critical density $\hat{\varrho}_c$ by $\epsilon\equiv
\hat{\varrho}/\hat{\varrho}_c-1$.  So, the $\epsilon<0$, $\epsilon=0$ and 
$\epsilon>0$ cases respectively correspond to the open, flat and closed 
universes.  The $\hat{t}$-derivative of $\epsilon$ is given by
\begin{equation}
\dot{\epsilon}=(1+\epsilon)\left({\dot{\hat{\varrho}}\over\hat{\varrho}}
-{\dot{\hat{\varrho}}_c\over\hat{\varrho}_c}\right).
\label{epsdot}
\end{equation}
We assume that the brane matter fields satisfy the equation of state of the 
form $\hat{\wp}=w\hat{\varrho}c^2_0$ with a constant $w$.  Then, making 
use of Eqs. (\ref{lnfrd1}-\ref{neceq}), we obtain
\begin{eqnarray}
{\dot{\hat{\varrho}}\over\hat{\varrho}}&=&-3{\dot{\hat{a}}_0\over\hat{a}_0}
(1+w)-{\dot{\hat{G}}_4\over\hat{G}_4}+{\dot{\varepsilon}\over\varepsilon}
\left(3w+{{\epsilon-1}\over{\epsilon+1}}-{{3\hat{\cal C}c^2_0}\over
{4\pi\hat{G}_4\hat{a}^4_0\hat{\varrho}}}\right),
\cr
{\dot{\hat{\varrho}}_c\over\hat{\varrho}_c}&=&-{\dot{\hat{a}}_0\over\hat{a}_0}
[2+(1+\epsilon)(1+3w)]-{\dot{\hat{G}}_4\over\hat{G}_4}+{\dot{\varepsilon}\over
\varepsilon}\left[(1+\epsilon)(1+3w)-2-{{3\hat{\cal C}c^2_0}\over{4\pi\hat{G}_4
\hat{a}^4_0\hat{\varrho}_c}}\right].
\label{dvarrhos}
\end{eqnarray}
Substituting these into Eq. (\ref{epsdot}), we obtain the following equation 
describing the time evolution of $\epsilon$:
\begin{eqnarray}
\dot{\epsilon}&=&(1+\epsilon)\epsilon{\dot{\hat{a}}_0\over\hat{a}_0}(1+3w)+
\left[2-(1+\epsilon)(1+3w)+{{3\hat{\cal C}c^2_0}\over{4\pi\hat{G}_4\hat{a}^4_0
\hat{\varrho}_c}}\right]\epsilon{\dot{\varepsilon}\over\varepsilon}
\cr
&=&(1+\epsilon)\epsilon{\dot{\hat{a}}_0\over\hat{a}_0}(1+3w)+\left[
{{2\left({\dot{\hat{a}}_0\over\hat{a}_0}-{\dot{\varepsilon}\over\varepsilon}
\right)^2}\over{\left({\dot{\hat{a}}_0\over\hat{a}_0}-{\dot{\varepsilon}
\over\varepsilon}\right)^2-{{\hat{\cal C}c^2_0\varepsilon^4}\over\hat{a}^4_0}}}
-(1+\epsilon)(1+3w)\right]\epsilon{\dot{\varepsilon}\over\varepsilon}.
\label{epdeq}
\end{eqnarray}
As in the VSL cosmologies, We assume that $\varepsilon$ varies rapidly enough 
such that $|\dot{\varepsilon}/\varepsilon|\gg\dot{\hat{a}}_0/\hat{a}_0$ during 
the early period of cosmic evolution.  Then, Eq. (\ref{epdeq}) is approximated 
to
\begin{equation}
\dot{\epsilon}\approx\left[{{2\left({\dot{\varepsilon}\over\varepsilon}
\right)^2}\over{\left({\dot{\varepsilon}\over\varepsilon}\right)^2-
{{\hat{\cal C}c^2_0\varepsilon^4}\over\hat{a}^4_0}}}-(1+\epsilon)(1+3w)
\right]\epsilon{\dot{\varepsilon}\over\varepsilon}.
\label{aprxepdep}
\end{equation}
Due to the upper limit on $\hat{\cal C}$ imposed by the nucleosynthesis 
constraint \cite{bdel} and the requirement that the dark radiation term 
should not give a large contribution to the current expansion of the 
universe, we can neglect the $\hat{\cal C}$-term in Eq. (\ref{aprxepdep}) 
compared to $(\dot{\varepsilon}/\varepsilon)^2$.  So, the terms in the 
square bracket of Eq. (\ref{aprxepdep}) can be approximated to $2-(1+\epsilon)
(1+3w)$.  The condition for the flat universe ($\epsilon=0$) to be a stable 
attractor is therefore $w\leq 1/3$.  This condition can be satisfied by the 
radiation brane matter ($w=1/3$) and the dust ($w=0$), so our varying-$e$ 
brane world cosmological model solves the flatness problem, if $\varepsilon$ 
varies rapidly enough.  

Next, we consider the resolution of the cosmological constant problem.  
Unlike the case of the standard cosmology, the mass density of the brane 
matter, satisfying $\hat{\varrho}=-\hat{\wp}/c^2_0$, is not directly 
related to the cosmological constant, but rather to the brane tension.  
Assuming the brane tension $\hat{\sigma}$ to have initially 
taken the fine-tunned value giving rise to the zero effective cosmological 
constant on the brane, we can regard the mass density 
$\hat{\varrho}_{\delta\hat{\sigma}}$ satisfying $\hat{\varrho}_{\delta
\hat{\sigma}}=-\hat{\wp}_{\delta\hat{\sigma}}/c^2_0$ as being due to the 
the correction $\delta\hat{\sigma}$ to the fine-tuned brane tension.  
Nonzero $\delta\hat{\sigma}$ gives rise to nonzero effective cosmological 
constant $\Lambda_{\rm eff}={{16\pi^2\hat{G}^2_5}\over{3c^8_0}}\delta
\hat{\sigma}^2$ in the four-dimensional universe on the brane.  We separate 
the contribution to $\hat{\varrho}$ into that $\hat{\varrho}_m$ from the 
ordinary brane matter and that $\hat{\varrho}_{\delta\hat{\sigma}}=\delta
\hat{\sigma}/c^2_0$ from the correction to the fine-tuned brane tension, 
i.e., $\hat{\varrho}=\hat{\varrho}_m+\hat{\varrho}_{\delta\hat{\sigma}}$.  
Then, the conservation equation (\ref{neceq}) is modified to
\begin{eqnarray}
\dot{\hat{\varrho}}_m+3{\dot{\hat{a}}_0\over\hat{a}_0}\left(\hat{\varrho}_m+
{\hat{\wp}_m\over c^2_0}\right)=-\dot{\hat{\varrho}}_{\delta\hat{\sigma}}
-{\dot{\hat{G}}_4\over\hat{G}_4}\hat{\varrho}+\left(3{\hat{\wp}_m\over c^2_0}
-\hat{\varrho}_m-4\hat{\varrho}_{\delta\hat{\sigma}}\right.
\ \ \ \ \ \ \ \ \ \ \ \ \ \ \ 
\cr
\left.+{{3\hat{k}c^2_0}\over{4\pi\hat{G}_4\hat{a}^2_0\varepsilon^2}}
-{{3\hat{\cal C}c^2_0}\over{4\pi\hat{G}_4\hat{a}^4_0\varepsilon^2}}\right)
{\dot{\varepsilon}\over\varepsilon}.
\label{spneceq}
\end{eqnarray}
To study the time evolution of the cosmological constant term in the Friedmann 
equations, we define $\epsilon_{\delta\hat{\sigma}}\equiv\hat{\varrho}_{\delta
\hat{\sigma}}/\hat{\varrho}_m$, whose $\hat{t}$-derivative is given by
\begin{equation}
\dot{\epsilon}_{\delta\hat{\sigma}}=\epsilon_{\delta\hat{\sigma}}\left(
{\dot{\hat{\varrho}}_{\delta\hat{\sigma}}\over\hat{\varrho}_{\delta\hat{
\sigma}}}-{\dot{\hat{\varrho}}_m\over\hat{\varrho}_m}\right).
\label{tdervdel}
\end{equation}
Making use of the effective Friedmann equations and the conservation equation 
and assuming the equation of state of the form $\hat{\wp}=w\hat{\varrho}
c^2_0$, we obtain
\begin{equation}
{\dot{\hat{\varrho}}_{\delta\hat{\sigma}}\over\hat{\varrho}_{\delta
\hat{\sigma}}}=-6{\dot{\varepsilon}\over\varepsilon},
\label{inre1}
\end{equation}
\begin{equation}
{\dot{\hat{\varrho}}_m\over\hat{\varrho}_m}=-3{\dot{\hat{a}}\over\hat{a}}
(1+w)+(3w-1){\dot{\varepsilon}\over\varepsilon}-2{\dot{\varepsilon}\over
\varepsilon}{\hat{\varrho}_c\over\hat{\varrho}_m}+2{\dot{\varepsilon}\over
\varepsilon}{{\hat{\varrho}+\hat{\varrho}_{\delta\hat{\sigma}}}\over
\hat{\varrho}_m}-{\hat{\varrho}\over\hat{\varrho}_m}{\dot{\hat{G}}_4\over
\hat{G}_4}-{{3\hat{\cal C}c^2_0}\over{4\pi\hat{G}_4\hat{a}^4_0}}
{\dot{\varepsilon}\over\varepsilon}.
\label{inre2}
\end{equation}
Substituting these into Eq. (\ref{tdervdel}), we obtain the following equation 
describing the time evolution of $\epsilon_{\delta\hat{\sigma}}$:
\begin{equation}
\dot{\epsilon}_{\delta\hat{\sigma}}=\epsilon_{\delta\hat{\sigma}}\left[3
\left({\dot{\hat{a}}_0\over\hat{a}_0}-{\dot{\varepsilon}\over\varepsilon}
\right)(1+w)+2{\dot{\varepsilon}\over\varepsilon}{{1+{\epsilon}_{\delta
\hat{\sigma}}}\over{1+\epsilon}}+\left({\dot{\hat{G}}_4\over\hat{G}_4}-4
{\dot{\varepsilon}\over\varepsilon}\right)(1+{\epsilon}_{\delta
\hat{\sigma}})+{{3\hat{\cal C}c^2_0}\over{4\pi\hat{G}_4\hat{a}^4_0}}
{\dot{\varepsilon}\over\varepsilon}\right].
\label{labeveq}
\end{equation}
If we assume $|\dot{\varepsilon}/\varepsilon|\gg\dot{\hat{a}}_0/
\hat{a}_0$, then Eq. (\ref{labeveq}) for the flat brane universe 
($\epsilon=0$) is approximated to
\begin{equation}
\dot{\epsilon}_{\delta\hat{\sigma}}=\epsilon_{\delta\hat{\sigma}}\left[
(1+\epsilon_{\delta\hat{\sigma}}){\dot{\hat{G}}_4\over\hat{G}_4}+
\left({{3\hat{\cal C}c^2_0}\over{4\pi\hat{G}_4\hat{a}^4_0}}-5-3w-2
\epsilon_{\delta\hat{\sigma}}\right){\dot{\varepsilon}\over\varepsilon}
\right].
\label{flatlabeveq}
\end{equation}
In order for $\epsilon_{\delta\hat{\sigma}}$ to be driven to zero (so that 
the correction $\delta\hat{\sigma}$ to the fine-tuned brane tension does not 
grow with time) as the brane universe expands, the term in the square bracket 
has to be negative.  With $\dot{\hat{G}}_4=0$, this can be achieved, if the 
following is satisfied:
\begin{equation}
{{3\hat{\cal C}c^2_0}\over{4\pi\hat{G}_4\hat{a}^4_0}}>5+3w+2
\epsilon_{\delta\hat{\sigma}}.
\label{cscnst}
\end{equation}
When $\dot{\hat{G}}_4\neq 0$, the condition for $\epsilon_{\delta
\hat{\sigma}}=0$ being a stable attractor becomes less stringent.


\begin{thebibliography} {99}
\small
\parskip=0pt plus 2pt

\bibitem{wfc} J.K. Webb, V.V. Flambaum, C.W. Churchill, M.J. Drinkwater and
J.D. Barrow, ``Evidence for time variation of the fine structure constant,''
Phys. Rev. Lett. {\bf 82} (1999) 884, astro-ph/9803165.
%%CITATION = ASTRO-PH 9803165;%% 

\bibitem{mur} M.T. Murphy {\it et al.}, ``Possible evidence for a variable 
fine structure constant from QSO absorption lines: motivations, analysis and 
results,'' Mon. Not. Roy. Astron. Soc. {\bf 327} (2001) 1208, 
astro-ph/0012419.
%%CITATION = ASTRO-PH 0012419;%%

\bibitem{web} J.K. Webb {\it et al.}, ``Further evidence for cosmological
evolution of the fine structure constant,'' Phys. Rev. Lett. {\bf 87} (2001)
091301, astro-ph/0012539.
%%CITATION = ASTRO-PH 0012539;%%

\bibitem{mof1} J.W. Moffat, ``Superluminary universe: A Possible solution to
the initial value problem in cosmology,'' Int. J. Mod. Phys. {\bf D2} (1993)
351, gr-qc/9211020. 
%%CITATION = GR-QC 9211020;%%

\bibitem{am} A. Albrecht and J. Magueijo, ``Time varying speed of light as 
a solution to cosmological puzzles,'' Phys. Rev. {\bf D59} (1999) 043516, 
astro-ph/9811018.
%%CITATION = ASTRO-PH 9811018;%%

\bibitem{gut} A.H. Guth, ``The inflationary universe: A possible solution to
the horizon and flatness problems,'' Phys. Rev. {\bf D23} (1981) 347. 
%%CITATION = PHRVA,D23,347;%%

\bibitem{lin} A.D. Linde, ``A new inflationary universe scenario: A possible
solution of The horizon, flatness, homogeneity, isotropy and primordial
monopole problems,'' Phys. Lett. {\bf B108} (1982) 389. 
%%CITATION = PHLTA,B108,389;%%

\bibitem{als} A. Albrecht and P.J. Steinhardt, ``Cosmology for grand unified
theories with radiatively induced symmetry breaking,'' Phys. Rev. Lett.
{\bf 48} (1982) 1220.
%%CITATION = PRLTA,48,1220;%%

\bibitem{bar1} J.D. Barrow, ``Cosmologies with varying light speed,''
Phys. Rev. {\bf D 59} (1999) 043515, astro-ph/9811022. 
%%CITATION = ASTRO-PH 9811022;%%

\bibitem{bar2} J.D. Barrow and J. Magueijo, ``Varying-alpha theories and
solutions to the cosmological problems,'' Phys. Lett. {\bf B443} (1998) 104,
astro-ph/9811072. 
%%CITATION = ASTRO-PH 9811072;%%

\bibitem{bar3} J.D. Barrow and J. Magueijo, ``Solutions to the quasi-flatness
and quasi-lambda Problems,'' Phys. Lett. {\bf B447} (1999) 246,
astro-ph/9811073. 
%%CITATION = ASTRO-PH 9811073;%%

\bibitem{mof2} J.W. Moffat, ``Varying light velocity as a 
solution to the problems in cosmology,'' astro-ph/9811390. 
%%CITATION = ASTRO-PH 9811390;%%
                               
\bibitem{cm1} M.A. Clayton and J.W. Moffat, ``Dynamical mechanism for varying
light velocity as a solution to cosmological problems,'' Phys. Lett.
{\bf B460} (1999) 263, astro-ph/9812481. 
%%CITATION = ASTRO-PH 9812481;%%
                               
\bibitem{bar4} J.D. Barrow and J. Magueijo, ``Solving the flatness and
quasi-flatness problems in Brans-Dicke  cosmologies with a varying light
speed,'' Class. Quant. Grav. {\bf 16} (1999) 1435, astro-ph/9901049.
%%CITATION = ASTRO-PH 9901049;%%

\bibitem{cm2} M.A. Clayton and J.W. Moffat, ``Scalar-tensor gravity theory
for dynamical light velocity,'' Phys. Lett. {\bf B477} (2000) 269,
gr-qc/9910112. 
%%CITATION = GR-QC 9910112;%%

\bibitem{blm} B.A. Bassett, S. Liberati, C. Molina-Paris and M. Visser, 
``Geometrodynamics of variable speed of light cosmologies,'' Phys. Rev. 
{\bf D62} (2000) 103518, astro-ph/0001441.
%%CITATION = ASTRO-PH 0001441;%%

\bibitem{cm3} M.A. Clayton and J.W. Moffat, ``Vector field mediated models
of dynamical light velocity,'' gr-qc/0003070. 
%%CITATION = GR-QC 0003070;%%

\bibitem{cm4} M.A. Clayton and J.W. Moffat, ``A scalar-tensor cosmological 
model with dynamical light velocity,'' Phys. Lett. {\bf B506} (2001) 177, 
gr-qc/0101126.
%%CITATION = GR-QC 0101126;%%

\bibitem{dru} I.T. Drummond, ``Variable light-cone theory of gravity,''
gr-qc/9908058.
%%CITATION = GR-QC 9908058;%%

\bibitem{bek} J.D. Bekenstein, ``Fine structure constant: Is it really a
constant?,'' Phys. Rev. {\bf D25} (1982) 1527.
%%CITATION = PHRVA,D25,1527;%%

\bibitem{sbm} H.B. Sandvik, J.D. Barrow and J. Magueijo, ``A simple
varying-alpha cosmology,'' astro-ph/0107512.
%%CITATION = ASTRO-PH 0107512;%%
 
\bibitem{youm} D. Youm, ``Brane world cosmologies with varying speed of 
light,'' Phys. Rev. {\bf D63} (2001) 125011, hep-th/0101228.
%%CITATION = HEP-TH 0101228;%%

\bibitem{youm1} D. Youm, ``Variable-speed-of-light cosmology from brane 
world scenario,'' Phys. Rev. {\bf D64} (2001) 085011, hep-th/0102194.
%%CITATION = HEP-TH 0102194;%%

\bibitem{youm2} D. Youm, ``A scalar tensor bimetric brane world cosmology,'' 
hep-th/0108073.
%%CITATION = HEP-TH 0108073;%%
                             
\bibitem{rs1} L. Randall and R. Sundrum, ``A large mass hierarchy from a 
small extra dimension,'' Phys. Rev. Lett. {\bf 83} (1999) 3370, hep-ph/9905221.
%%CITATION = HEP-PH 9905221;%%

\bibitem{rs2} L. Randall and R. Sundrum, ``An alternative to 
compactification,'' Phys. Rev. Lett. {\bf 83} (1999) 4690, hep-th/9906064.
%%CITATION = HEP-TH 9906064;%%

\bibitem{kal1} G. Kalbermann and H. Halevi, ``Nearness through an extra
dimension,'' gr-qc/9810083. 
%%CITATION = GR-QC 9810083;%%

\bibitem{kir} E. Kiritsis, ``Supergravity, D-brane probes and thermal super
Yang-Mills:  A comparison,'' JHEP{\bf 9910} (1999) 010, hep-th/9906206. 
%%CITATION = HEP-TH 9906206;%%

\bibitem{kal2} G. Kalbermann, ``Communication through an extra dimension,''
Int. J. Mod. Phys. {\bf A15} (2000) 3197, gr-qc/9910063. 
%%CITATION = GR-QC 9910063;%%

\bibitem{chu} D.J. Chung and K. Freese, ``Can geodesics in extra dimensions
solve the cosmological horizon  problem?,'' Phys. Rev. {\bf D 62} (2000)
063513, hep-ph/9910235. 
%%CITATION = HEP-PH 9910235;%%

\bibitem{ale} S.H. Alexander, ``On the varying speed of light in a
brane-induced FRW universe,'' JHEP{\bf 0011} (2000) 017, hep-th/9912037. 
%%CITATION = HEP-TH 9912037;%%

\bibitem{ish} H. Ishihara, ``Causality of the brane universe,'' Phys. Rev.
Lett. {\bf 86} (2001) 381, gr-qc/0007070. 
%%CITATION = GR-QC 0007070;%%

\bibitem{ckr} D.J. Chung, E.W. Kolb and A. Riotto, ``Extra dimensions present
a new flatness problem,'' hep-ph/0008126. 
%%CITATION = HEP-PH 0008126;%%

\bibitem{csa} C. Csaki, J. Erlich and C. Grojean, ``Gravitational Lorentz 
violations and adjustment of the cosmological  constant in asymmetrically 
warped spacetimes,'' Nucl. Phys. {\bf B604} (2001) 312, hep-th/0012143.
%%CITATION = HEP-TH 0012143;%%

\bibitem{youm3} D. Youm, ``A varying-alpha brane world cosmology,'' 
hep-th/0108237.
%%CITATION = HEP-TH 0108237;%%

\bibitem{bdl} P. Binetruy, C. Deffayet and D. Langlois, ``Non-conventional 
cosmology from a brane-universe,'' Nucl. Phys. {\bf B565} (2000) 269, 
hep-th/9905012.
%%CITATION = HEP-TH 9905012;%%

\bibitem{bdel} P. Binetruy, C. Deffayet, U. Ellwanger and D. Langlois,
``Brane cosmological evolution in a bulk with cosmological constant,''
Phys. Lett. {\bf B477} (2000) 285, hep-th/9910219. 
%%CITATION = HEP-TH 9910219;%%

\bibitem{ftw} E.E. Flanagan, S.H. Tye and I. Wasserman, ``Cosmological
expansion in the Randall-Sundrum brane world scenario,'' Phys. Rev. {\bf D62}
(2000) 044039, hep-ph/9910498.
%%CITATION = HEP-PH 9910498;%%

\bibitem{muk} S. Mukohyama, ``Brane-world solutions, standard cosmology,
and dark radiation,'' Phys. Lett. {\bf B473} (2000) 241, hep-th/9911165. 
%%CITATION = HEP-TH 9911165;%%

\bibitem{ida} D. Ida, ``Brane-world cosmology,'' JHEP {\bf 0009} (2000) 014, 
gr-qc/9912002.
%%CITATION = GR-QC 9912002;%%

\bibitem{cgk} C. Csaki, M. Graesser, C. Kolda and J. Terning, ``Cosmology
of one extra dimension with localized gravity,'' Phys. Lett. {\bf B462}
(1999) 34, hep-ph/9906513. 
%%CITATION = HEP-PH 9906513;%%

\bibitem{cgs} J.M. Cline, C. Grojean and G. Servant, ``Cosmological expansion
in the presence of extra dimensions,'' Phys. Rev. Lett. {\bf 83} (1999) 4245,
hep-ph/9906523.
%%CITATION = HEP-PH 9906523;%%

\end{thebibliography}
\end{document}